\documentclass[12pt]{article}
\usepackage{cite}
\usepackage{amsmath,amsfonts,amssymb}
\usepackage{graphicx,color}
\usepackage{epsfig,epsf}
\usepackage{bm}

\usepackage{relsize}

\usepackage[bookmarks, breaklinks, colorlinks,urlcolor=black, citecolor=red, 
linkcolor=blue]{hyperref}

\usepackage{url}
 \usepackage{cancel}

\def\dsp{\displaystyle}

\def\gev{\ensuremath{\mathrm{Ge\kern -0.1em V}}}

\begin{document}

\renewcommand*{\thefootnote}{\fnsymbol{footnote}}

%%%%%%%%%%%%%%%%%%%%%%%%%%%%%%%%%%%%%%%%%%%%%%%%%%%%%%%
\begin{center}
 {\Large\bf{Fermionic dark matter in leptoquark portal}}
 \\[6mm]
 { Rusa Mandal\,\footnote{Electronic address: Rusa.Mandal@ific.uv.es}}
\\[3mm]

{\small\em The Institute of Mathematical Sciences, HBNI, Taramani, Chennai 600113, India \\ and \\ IFIC, Universitat de Val$\grave{e}$ncia-CSIC, Apt. Correus 22085, E-46071 Val$\grave{e}$ncia, Spain}
 \end{center}
%%%%%%%%%%%%%%%%%%%%%%%%%%%%%%%%%%%%%%%%%%%%%%%%%%%%%%%

\date{\today}
%\vspace{5mm}

\begin{abstract}

We investigate a beyond standard model (SM) portal scenario for dark matter (DM) particle with leptoquark being the mediator field. Leptoquark, a colored particle having both baryon and lepton number, allows the DM to interact with the SM fields via renormalizable interaction.
By focusing on a vector leptoquark portal with Majorana fermion DM candidate, we find the only unknown coupling in the model is sensitive to all three main features of a DM model namely, relic density, direct detection as well as indirect detection, while being consistent with collider data. We explore the parameter space of the portal with minimum of its field content and find that AMS-02 data for antiproton flux imposes stringent bound till date and excludes the DM mass up to $400\,$GeV. The LUX 2016 data for DM-neutron scattering cross section allows the region compatible with relic density, however the future sensitivity of LUX-ZEPLIN experiment can probe the model up to its perturbative limit.

\end{abstract}

%%%%%%%%%%%%%%%%%%%%%%%%%%%%%%%%%%%%%%%%%%%%%%%%%%%%%%%%%%%%%%%%%%%%%%%%%%%%%%

\section{Introduction}

The cosmological evidences of  missing mass hypothesized as dark matter (DM) support its gravitational interaction. No other type of interactions for DM is found till date. From particle physics perspective various different interactions of the DM candidate are speculated. Within the standard model (SM), while the most promising DM candidate i.e., neutrinos, turn out incapable of being a cold DM due to very light mass and non-relativistic nature, we are forced to extend the SM beyond its particle content. Since the coupling of the DM candidate to the SM particles are unknown, it is worthwhile to explore all possible extensions of the SM with manifest non-gravitational interactions.

From the last three decades, plethora of candidates have been imagined as a DM particle in literature. Among them one of the most popular choice is a weakly interacting and massive particle (WIMP) whose mass lies in the GeV to TeV range with typically weak interactions. A WIMP pair annihilation to the SM particles provides a natural mechanism to produce the WIMP at the early Universe and can also explain the observed DM density in current Universe. Now, in the case of SM gauge singlet DM candidates, the mediation of dark interaction to the SM fields through renormalizable interactions are known as ``portal". Within the SM, there are three such portals namely, Higgs portal~\cite{Silveira:1985rk}, gauge boson portal~\cite{Galison:1983pa} and neutrino portal~\cite{Minkowski:1977sc}. The portals can also comprise beyond standard model (BSM) particles e.g. $Z^\prime$ portal~\cite{Dudas:2009uq}.

In this work we study the minimal model with leptoquark portal. Leptoquarks, a spin zero or spin one particle, can turn a lepton to quark and vice-verse. In several extensions of the SM, which treat the leptons and quarks in the same basis, like $SU(5)$~\cite{Georgi:1974sy}, $SU(4)_C\times SU(2)_L\times SU(2)_R$~\cite{Pati:1974yy}, or $SO(10)$~\cite{Georgi:1974my} and also theories with composite model~\cite{Dimopoulos:1979es} and technicolor model~\cite{Farhi:1980xs} can contain such particles. Leptoquarks carry both baryon and lepton numbers simultaneously. Depending on the SM gauge quantum numbers of the leptoquark, renormalizable interaction terms consisting of SM fields, leptoquark and a SM singlet fermion can automatically be included in the theories where the singlet fermion may serve as a cosmologically stable DM candidate.

Based on the spin of the leptoquark, the portal can be terms as scalar leptoquark  portal or vector leptoquark portal. Few previous studies on scalar portal models can be found in Refs.~\cite{Arcadi:2017kky,Allahverdi:2017edd,Agrawal:2011ze,Fornal:2018eol}. In most of these analyses, the main focus is at the low DM mass region. Another similar framework with scalar mediator is also studied in Ref.~\cite{Garny:2018icg}  where apart from the WIMP paradigm, a different region in the parameter space is also explored based on the mechanism of conversion-driven freeze-out. In this paper, we 
concentrate on the vector leptoquark portal which is less studied compared to the scalar portal. Being a vector particle, this scenario has greater chances to be probed in direct as well as indirect detection experiment due to the larger cross section.  Our choice of the particular leptoquark is the very first leptoquark which has been described in literature in Pati-Salam model~\cite{Pati:1974yy}. Here we only study a simplified version  which is assumed to describe the physics at the phenomenologically relevant scales of a UV complete theory to good approximation. As the leptoquark under discussion is a vector particle, it should be gauged under some symmetry group in the UV complete model. The spontaneous breaking of the symmetry will render mass for the leptoquark. Some attempts to write a UV model can be found in Refs.~\cite{DiLuzio:2017vat,Assad:2017iib}. As we see in the next section, to stabilize the DM particle we need to impose some discrete symmetry which can be a remnant of some larger symmetry of the UV theory. Recently, the vector leptoquark has also received significant impetus for having the possibility in explaining the anomalies observed in charged-current as well as neutral current transitions of $B$ mesons~\cite{Barbieri:2015yvd,DiLuzio:2017chi,Choudhury:2017qyt,Blanke:2018sro,Crivellin:2018yvo}. 

The leptoquark portal models have the detectability through direct detection experiments. In the case of dirac fermion DM particle, both spin-independent (SI) and spin-dependent (SD) scattering cross sections exist and hence provide more opportunity to probe such scenario compared to a Majorana fermion DM where only spin-dependent cross section survives. At recent time, the direct detection experiments especially XENON1T~\cite{Aprile:2018dbl}, LUX~\cite{Akerib:2016vxi} impose very stringent bounds on the DM-nuclei cross section. In this work, we discuss the case with a Majorana fermion DM particle.

Indirect detection techniques to detect the cosmic ray fluxes through the dedicated detectors can receive signals from leptoquark portal models as well. Depending on the final states to which the DM pair annihilates dominantly, Fermi Large Area Telescope (LAT) and Alpha Magnetic Spectrometer (AMS) measurements can be very relevant and thus providing yet another way to probe this type of models.

The rest of the paper is organized as follows. In Sec.~\ref{sec:portal}, we discuss the possible leptoquark portals which can arise from the minimal field content of the SM and then focus on an explicit vector leptoquark portal model. Section \ref{sec:relic} contains the details of relic abundance calculation and in Sec.~\ref{sec:DD}, the DM-nuclei cross section for direct detection experiments are described. Combining the findings of the previous sections, we illustrate the results in Sec.~\ref{sec:results} with the discussion on recent collider bounds and indirect detection limits. Finally, we summarize in Sec.~\ref{sec:sum}.

%%%%%%%%%%%%%%%%%%%%%%%%%%%%
\section{Leptoquark portal}
\label{sec:portal}
%%%%%%%%%%%%%%%%%%%%%%%%%%%%

In this section we briefly discuss the possible choices of leptoquark portals. By suppressing the color and generation indicies, we quote such possible interaction terms with the SM quantum numbers and spin of the mediator leptoquark, as the following.
%
%\begin{table}[h]
\begin{center}
	\begin{tabular}{ccc}
		% \hline
		Interaction & $\left(SU(3)_C,\,SU(2)_L,\,U(1)_Y\right)$ 
		 & Spin\cr 	\hline  \rule{0pt}{4ex}
		$\bar{d}_R^C\, X\, \psi$ & $({\bf\bar{3}},\,{\bf1},\,1/3)$ & 0 ~ \cr  \rule{0pt}{4ex}
		$\bar{u}_R^C\, X\, \psi$ & $({\bf \bar{3}},\,{\bf1},\, -2/3)$ & 0 ~ \cr \rule{0pt}{4ex}
		$\bar{Q}_L X\, \psi$ & $({\bf 3},\,{\bf2},\,1/6)$ & 0 ~ \cr \rule{0pt}{4ex}
		$\bar{Q}_L^C\,\gamma^\mu X_\mu\, \psi$ & $({\bf \bar{3}},\,{\bf2},\,-1/6)$ & 1 ~ \cr \rule{0pt}{4ex}
		$\bar{u}_R\,\gamma^\mu X_\mu\, \psi$ & $({\bf 3},\,{\bf1},\,2/3)$ & 1 ~ \cr \rule{0pt}{4ex}
		$\bar{d}_R\,\gamma^\mu X_\mu\, \psi$ & $({\bf 3},\,{\bf1},\,-1/3)$ & 1 ~ \cr 
		%\hline
	\end{tabular}
%	\caption{The list of leptoquark portal interactions with the SM quantum numbers and spin of the leptoquark are shown.}
	\end{center}
%	\end{table}

%
Here $Q_L$ and $u_R$, $d_R$ are the SM $SU(2)_L$ quark doublets and right-handed singlets, respectively. The boson $X_{(\mu)}$ denotes the scalar (vector) leptoquark and $\psi$ is a SM singlet fermion, the would be DM candidate.

For first four cases, the interaction of the leptoquark with other SM fields induce baryon number violating processes~\cite{Dorsner:2016wpm} and thus generally give rise to proton decay. Hence we avoid such types from our considerations. It is should be noted that in the case of Dirac DM particle, the SI DM-nuclei cross section measurements by XENON1T~\cite{Aprile:2018dbl} and LUX~\cite{Akerib:2016vxi} experiments almost completely exclude the parameter space at its minimal content. For instance for the last two cases, if $\psi$ is an $\mathcal{O}$\,(100 GeV) Dirac DM particle, for $\mathcal{O}$\,(TeV) mediator mass, the coupling of the interaction term $>0.05$ is excluded at 90\% C.L. by XENON1T 2018 data~\cite{Aprile:2018dbl}, whereas the required coupling to satisfy observed relic density is one order higher. Hence we prefer to consider Majorana DM candidate for the analysis of this work.  Now we see in the next section, that the thermal average annihilation cross section of Majorana DM pair is proportional to the square of the mass of final states quarks and thus in the case of DM pair annihilating to down-type quark anti-quark pair, the annihilation rate is insufficient to produce the observed relic density within the perturbative limit of the interaction strength. Hence the only reasonable choice among the six portals shown above reduces to the fifth portal namely a vector leptoquark with $({\bf 3},\,1,\,2/3)$ quantum numbers under the SM gauge group. In rest of the paper we explore this particular portal elaborately.

The Lagrangian of such construct can be written as
\begin{align}
\label{eq:Lag}
\hspace*{-0.27cm}\mathcal{L} \subset  -\frac{1}{2} U_{\mu\nu}^\dagger U^{\mu\nu}\! + m_U^2\, U_\mu^\dagger U^\mu - \frac{1}{2} m_\chi \chi \chi - y_L \bar{Q}_L\gamma_\mu U^\mu L_L - y_R\, \bar{d}_R \gamma_\mu U^\mu e_R- y_\chi  \bar{u}_R\gamma_\mu U^\mu \chi + {\rm h.c.}\,,
\end{align}
where $ U_{\mu\nu}= D_\nu\, U_\mu - D_\mu\, U_\nu$ with $D_\mu=\partial_\mu - i g_s \dsp\frac{\lambda^a}{2} G_\mu^a- i g^\prime\, \dsp\frac{2}{3} B_\mu $.
Here $L_L$ and $e_R$  are the SM $SU(2)_L$ lepton doublets and right-handed singlets, respectively. The new fields, $U$ is a vector leptoquark with charges under the SM as $({\bf 3},\,1,\,2/3)$ whereas $\chi$ is a Majorana fermion being singlet under the SM gauge group. If $m_\chi<m_{U}$, the two-body decay of $\chi$ is forbidden at tree level. However, the interaction of $U$ with the SM fields written in Eq.~\eqref{eq:Lag} will induce tree level three body decay and one-loop induced decay of $\chi$. Such decays can be avoided if we introduce an extra symmetry, namely a $Z_2$ symmetry and assume only the leptoquark $U$ and $\chi$ are odd under it. In such case, the fermion $\chi$ can serve as a cosmological stable DM candidate. There have also been studies~\cite{Garny:2012vt,Arcadi:2013aba,Arcadi:2014dca} where by relaxing the extra symmetry arguments, the constraint on DM lifetime is imposed from observed cosmological signal while being consistent with the thermal freeze out condition. In this case, the couplings $y_{L,R}$ turn out to be highly suppressed for $\mathcal{O}(1)$ values of $y_\chi$. Hence for the analysis of this paper, we assume $y_{L,R}$ to be vanishing and only last term of Eq.~\eqref{eq:Lag} is the relevant interaction which connects the visible sector to the dark sector ($\chi$). 

In writing Eq.~\eqref{eq:Lag}, we have assumed minimal coupling scenario i.e., the interaction term $i g_s U_\mu^\dagger \dsp\frac{\lambda^a}{2} U_\nu G_{\mu\nu}^a$ is absent.
It should be noted that the coupling of DM particle $\chi$ to the three generations of up-type right-handed quarks can in general be different, however for simplicity, in this analysis we %adopt minimal flavor violation (MFV) scenario and 
assume them to be identical for all three generations. Similarly, the Yukawa type couplings $y_{L,R}$ are general matrices with off-diagonal terms which are stringently constrained by low energy processes like decays of Kaons and $B$ mesons. Nevertheless, the purpose of this work does not depend on the texture of such couplings and we refer to previous works in literature~\cite{Valencia:1994cj,Davidson:1993qk} for this discussion.

\subsection{Relic density}
\label{sec:relic}

\begin{figure}[h]
	\begin{center}
		\includegraphics[width=0.14\linewidth]{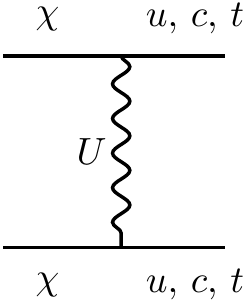}
		\caption{The Feynman diagram for the DM particle $\chi$ annihilation to up-type quark anti-quark pair. }\label{dia:relic}
	\end{center}
\end{figure}

It can be seen from Eq.~\eqref{eq:Lag} that the DM candidate $\chi$ can annihilate into the SM up-type quark anti-quark pair via a $t$-channel exchange of $U$ (shown in Fig.~\ref{dia:relic}). The thermal average annihilation cross section is given by,
\begin{align}
\label{eq:sigv}
\langle\sigma v \rangle = \frac{3\,y_\chi^4 m_q^2}{8 \pi \left( m_\chi^2 +m_{U}^2 -m_q^2 \right)^2} \left(1- \frac{m_q^2}{m_\chi^2}\right)^{1/2},
\end{align}
\noindent
where $m_q$ is the mass of up-type quark. It is apparent that for $m_\chi>m_t$, $\chi\chi \to t\bar{t}$ is the most dominant annihilation mode. 

For  $m_\chi<m_t$, with $\mathcal{O}$(1) coupling, the annihilation channels to $\chi\chi \to u\bar{u},~c\bar{c}$  are insufficient to explain the observed relic density at present Universe. In such parameter space the processes like $\chi\chi\to gg,~Wtb$, can also be important. In context of a scalar leptoquark portal with $\mathcal{O}\,$(GeV) DM candidate, non-thermal production of DM can also be an alternative way to produce the relic density as discussed in Ref.~\cite{Arcadi:2017kky}.

The model under consideration also poses co-annihilation channels such as $\chi U\to tg$ through a $t$-channel $U$ exchange. This process is significant only when the DM and the mediator are very close in masses. Another interesting co-annihilation mode, through a $s$-channel top quark, is $\chi U\to Wb$ and will only be efficient near the top quark resonance.

The observed relic abundance by Planck data $\Omega h^2 = 0.1199 \pm 0.0027$ ~\cite{Ade:2015xua} can be achieved by the thermal freeze-out condition,
$\langle \sigma v \rangle \approx 2 \times 10^{-9}\, \mbox{GeV}^{-2}$ and we explore the region in the next section.

\subsection{Direct detection}
\label{sec:DD}

The direct detection signal for this type of model can arise from the diagram shown in Fig.~\ref{dia:relic}. After integrating out the vector leptoquark and keeping terms only at leading order in $\partial^2/m_{U}^2$, the effective Lagrangian can be written as 
\begin{align}
\mathcal{L}_{\rm eff} \simeq -\frac{y_\chi^2}{4\left(m_{U}^2- m_\chi^2\right)}\, \bar{\chi} \left(1-\gamma_5\right) \gamma_\mu u\,\, \bar{u}\left(1-\gamma_5\right) \gamma^\mu\,  \chi\,.
\end{align}
Due to the fact that for a Majorana fermion $\bar{\chi}\gamma_\mu\chi=0$, after performing the Fierz transformation we are left only with the spin-dependent interaction given by
\begin{align}
\label{eq:Leff}
\mathcal{L}_{\rm eff} \simeq d_u\, \bar{u}  \gamma_\mu \gamma_5 u\,\bar{\chi}  \gamma_\mu \gamma_5 \chi \,;~~d_u\equiv -\frac{y_\chi^2}{4\left(m_{U}^2- m_\chi^2\right)}.
\end{align}
The DM-nucleon scattering cross section is expressed as~\cite{Agrawal:2010fh}
\begin{align}
\label{eq:sigD}
\sigma_{\rm SD}= \frac{16 m_\chi^2 m_N^2}{\pi \left(m_\chi + m_N\right)^2} \, d_u^2  {\Delta_u^N}^2\,J_N \left(J_N+1\right),
\end{align}
where $m_N\simeq1\,$GeV and $J_N=1/2$ are the mass and spin of a nucleon, respectively. The factor $\Delta_u^N$ denotes the spin fraction carried by a $u$-quark inside a nucleon and the estimates are $\Delta_u^p=0.78 \pm 0.02 $ for proton and $\Delta_u^n=-0.48 \pm 0.02 $ for neutron~\cite{Mallot:1999qb,Ellis:2000ds}.

\section{Results}
\label{sec:results}

The discussion in the preceding sections lead us to explore the parameter space of the model. The thermal averaged annihilation cross section (in Eq.~\eqref{eq:sigv}) and the DM-nucleon scattering cross sections (in Eq.~\eqref{eq:sigD}) are the two key predictions of the portal under consideration. It can be seen from Eq.~\eqref{eq:sigv} and Eq.~\eqref{eq:sigD}, both the expressions depend on three parameters, namely $m_\chi$, $m_{U}$ and $y_\chi$. As mentioned earlier for $m_\chi<m_t$, the annihilation cross section of the DM pair to lighter up-type quark antiquark is inadequate to produce the observed relic density. Thus we consider $m_\chi>m_t$ and vary up to $\mathcal{O}$(TeV). 

{\em  Collider bounds:} The searches at the LHC constrain both the mediator and DM particle in following ways. The leptoquarks can directly be searched for via pair and/or single production at the colliders. The signatures for the considered leptoquark portal model are $U\bar{U}\to t\bar{t}\chi\chi,~jj\chi\chi$ topologies. There exists other interesting final states arising from the first two terms of Eq.~\eqref{eq:Lag}. However as mentioned in Sec.~\ref{sec:portal}, these couplings are highly suppressed for the stability of the DM and thus are difficult to probe at current colliders~\cite{Arcadi:2014tsa}. Both ATLAS and CMS have searched for the decay of leptoquarks into three generations of quarks with missing energy coming either from the SM neutrinos or a DM particle. By rescaling a ATLAS search with 8\,TeV data~\cite{Aad:2015caa} for pair-produced third generation scalar leptoquark in the $t\bar{t}\nu_\tau\bar{\nu}_\tau$ channel,  $m_U < 770$\,GeV is excluded at $95\%$ confidence level with $\mathcal{B}(U\bar{U}\to t\bar{t}\nu_\tau\bar{\nu}_\tau)=0.25$~\cite{Barbieri:2015yvd}. In a search by CMS with 13\,TeV data~\cite{Sirunyan:2017yrk}, $m_U \lesssim 1$\,TeV is excluded while imposing the bounds on vector leptoquarks~\cite{DiLuzio:2017chi}. A latest results from CMS with 35.9\,fb$^{-1}$ data \cite{Sirunyan:2018kzh} imposes most stringent bound $m_U > 1.5$\,TeV to date by considering 100\% branching fraction to $t\nu$. As we assume the coupling of vector leptoquark $U$ to up-type quarks and DM particle is same for all three generations, the maximum branching fraction $\mathcal{B}(U\to \bar{t}\chi)$ is around 30\% and thus the limit on vector leptoquark mass can be relaxed further in our model. To be consistent within the collider limits of $1\sigma$ uncertainty we chose $m_U > 1$\,TeV for the analysis of this work.

\noindent
Another set of important collider signatures for the model under consideration is the search for monojet + $\cancel E_T$ and monophoton + $\cancel E_T$  at the LHC. While the later one being an electroweak process, is suppressed, the prior mode can be relevant to look for at the LHC~\cite{Khachatryan:2014rra,Aad:2015zva,Aaboud:2016tnv,CMS:2017tbk,Aaboud:2017phn}. The $t$-channel exchange of the vector leptoquark gives rise to final states with DM particles and jets. The most recent limit on monojet + $\cancel E_T$ channel is from ATLAS collaboration at 13\,TeV center-of-mass energy data with an integrated luminosity of $36.1$\,fb$^{-1}$. To generate the parton level cross section for the process $pp\to \chi\chi j$, we use {\tt MadGraph5}~\cite{Alwall:2014hca} where the model files are created by ${\tt FeynRules}$~\cite{Alloul:2013bka}. We use {\tt NNPDF23LO}~\cite{Ball:2012cx} parton distribution function (PDF) with 5 flavor quarks in initial states. The basic cuts used are the following $\cancel E_T > 250$ GeV, a leading jet with transverse momentum $p_T > 250$ GeV and pseudorapidity $|\eta| < 2.4$. Due to large
parton distribution probability of the gluon as compared to the quark or anti-quark in the proton, the $qg\to \chi \chi q$ process dominates. We find the region satisfying relic density with $y_\chi=1$ for low DM mass $m_\chi\lesssim200\,$GeV is excluded at $95\%$ confidence level. A similar observation was made in Ref.~\cite{An:2013xka} in context of a scalar leptoquark mediator with Majorana DM candidate using 8\,TeV data from CMS collaboration~\cite{Khachatryan:2014rra}. It should be noted that to satisfy the relic density with higher value of $y_\chi$, the mediator mass $m_U$ should also increase and in that case the cross section for $pp\to \chi \chi j$ decreases rendering weaker bounds from collider searches.

{\em Indirect detection bounds:} Indirect detection with gamma rays and antiprotons plays an important role to test the DM self-annihilating nature predicted by thermal freeze-out condition. The latest AMS-02~\cite{Aguilar:2016kjl} cosmic ray antiproton data imposes strong bound on the scenario we consider here. Fermi-LAT observations of
dwarf spheroidal galaxies also limits the annihilation cross sections however are weaker compared to the AMS-02 data. In Ref.~\cite{Cuoco:2017iax}, the limits on DM pair annihilation cross section into different SM fields have been obtained by using AMS-02 measurements of antiproton flux. For the case in our consideration, the dominant annihilation mode for the DM pair is to $t\bar{t}$ and we use the bound in our final results.

\begin{figure*}[!t]
	\begin{center}
		\includegraphics[width=0.6\linewidth]{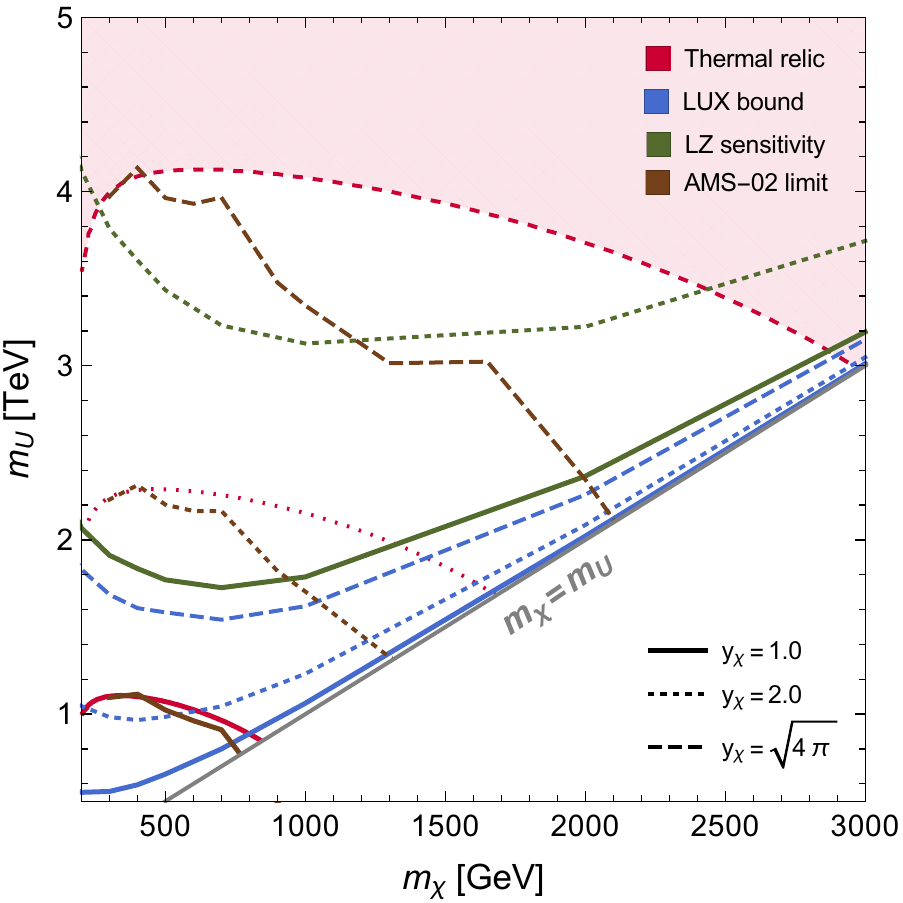}
		\caption{The allowed region the $m_\chi-m_U$ plane for different choices of coupling $\lambda_\chi=1$ (solid), 2 (dotted) and perturbative limit $\sqrt{4\pi}$ (dashed) are shown. The red curves satisfy the observed relic density~\cite{Ade:2015xua}. The blue and green curves represent current limit from spin-dependent DM-neutron scattering cross section measurement by LUX~\cite{Akerib:2016lao} and future sensitivity of LZ experiment~\cite{Akerib:2015cja}, respectively for the corresponding values of $y_\chi$. The region below these direct detection bounds are excluded at $90\%$ confidence level. The brown curves depict the indirect detection bound from AMS-02 measurements~\cite{Aguilar:2016kjl,Cuoco:2017iax} which excludes $m_\chi\leq 400\,$GeV at $95\%$ significance level. The rest of the parameter space is allowed by the current limits from both direct and indirect detection experiments however will completely be probed with future sensitivity of LZ experiment. }\label{fig:all}
	\end{center}
\end{figure*}

Combining the above discussions, we highlight the allowed region the $m_\chi-m_U$ plane for different choices of coupling $y_\chi$. The red curves satisfy the observed relic abundance by Planck data $\Omega h^2 = 0.1199 \pm 0.0027$~\cite{Ade:2015xua} where the texture solid, dot and dash denote $y_\chi=1,~2$ and $\sqrt{4\pi}$, respectively. The red shaded region is forbidden by perturbativity limit. The blue and green curves present current limit from SD DM-neutron scattering cross section measurement by LUX~\cite{Akerib:2016lao} and future sensitivity of LUX-ZEPLIN (LZ) experiment~\cite{Akerib:2015cja}, respectively, for the corresponding values of $y_\chi$. The region below these direct detection bounds are excluded at $90\%$ confidence level. It can be seen that for the current limit from LUX data, starting from $y_\chi=1$, all higher values are allowed and we have the entire parameter space compatible with relic abundance and direct detection limit. However, the future sensitivity of LZ will rule-out all the parameter space up to the perturbative limit of $y_\chi$. We also impose the bounds on annihilation cross section from AMS-02~\cite{Aguilar:2016kjl} measurements of antiproton flux depicted in brown curves for the respective $y_\chi$ values. This is currently the most stringent constraint on the model under consideration and for $m_\chi\leq 400\,$GeV, the region satisfying thermal freeze-out condition is excluded at $95\%$ confidence level.

We have also checked with the DM-proton scattering cross section limits measured in LUX experiment~\cite{Akerib:2015cja} which provides significantly less constrain on the parameter space compared to the DM-neutron case, as the measurements are one order less stringent.

We would like to mention that the portal discussed here can contribute to a low energy process namely $D_0-\bar{D}_0$ mixing via a box diagram mediated by the vector leptoquark $U$ and DM particle $\chi$. Such amplitude is quadratically divergent and thus one needs to introduce a cut-off scale $\Lambda$. For an $\mathcal{O}$(TeV) $\Lambda$, the measured limit on $|m_{D_1^0}- m_{D_1^0}|$ \cite{Patrignani:2016xqp} imposes strong constraint on coupling $y_\chi$ of the DM with $u$ and/or $c$ quark. Such calculation is regularisation dependent and can result in different bounds on $y_\chi$.  This limit is, however, in apparent conflict with the $\mathcal{O}$(1) values required for direct detection. We infer that one way to avoid such discordance is by invoking cancellation between fields due to yet-to-be-discovered symmetry of the full UV theory, analogous to GIM cancellation.

\section{Summary}
\label{sec:sum}

In this work we discuss the class of DM models dealing with leptoquark portal. The dark sector interacts with the SM fields via renormalizable interactions consisting of leptoquark, quark and the DM particle. By briefly reviewing the current status of the different models in leptoquark portal, we concentrate on a vector leptoquark portal with the leptoquark having charges under the SM group as $({\bf 3},\,1,\,2/3)$.

We consider a Majorana fermion DM candidate and assume the DM couples to all generations of quarks with equal coupling. This choice provides minimal scenario with only three free parameters in the model namely, the coupling $y_\chi$, mediator mass $m_U$ and the DM mass $m_\chi$. This simplistic model then is tested with three key features of a DM model as following.

The collider searches for leptoquarks impose bound on the leptoquark mass $\geq1\,$TeV which enforce us to focus mostly the high mass regions. The DM pair annihilation to $t\bar{t}$ can produce the observed relic density with $\mathcal{O}(1)$ values of the coupling $y_\chi$. However, for $m_\chi<m_t$, the observed relic abundance is not achievable within the perturbative regime of the couplings. The same coupling $y_\chi$ is also responsible for the DM-nucleon scattering cross section measured at direct detection experiments. In Fig.~\ref{fig:all}, we show the combined bound from LUX 2016 data on DM-neutron SD cross section measurements. The parameter space compatible with relic density is allowed by such measurements however, the proposed sensitivity of LZ experiment can rule-out the region up to the perturbative limit. The bound from DM-proton SD cross section data from LUX is less stringent than the case of DM-neutron measurements.

The interesting observation from indirect detection experiment namely, AMS-02 data for antiproton flux imposes stringent bound in the model under consideration. We illustrate in Fig.~\ref{fig:all} that the DM mass $\lesssim 400\,$GeV satisfying the relic constraint is excluded by such measurements at $95\%$ confidence level.

\vspace*{-0.2cm}
\subsection*{Acknowledgment}
\vspace*{-0.2cm}
We are thankful to Rahul Sinha, Eung Jin Chun and Priyotosh Bandyopadhyay for very important and insightful suggestions. We thank Celine Degrande and Narayan Rana for the help in ${\tt FeynRules}$. This work has been supported in part by Grants No. FPA2014-
53631-C2-1-P, FPA2017-84445-P and SEV-2014-0398 (AEI/ERDF, EU) and by PROMETEO/2017/053.


\begin{thebibliography}{99}

%\cite{Silveira:1985rk,McDonald:1993ex,He:2016mls}
\bibitem{Silveira:1985rk}
  V.~Silveira and A.~Zee,
  %``Scalar Phantoms,''
  Phys.\ Lett.\  {\bf 161B} (1985) 136.
%  doi:10.1016/0370-2693(85)90624-0
  %%CITATION = doi:10.1016/0370-2693(85)90624-0;%%
  %435 citations counted in INSPIRE as of 04 Sep 2017

%\cite{Galison:1983pa}
\bibitem{Galison:1983pa} 
  P.~Galison and A.~Manohar,
  %``TWO Z's OR NOT TWO Z's?,''
  Phys.\ Lett.\  {\bf 136B}, 279 (1984).
%  doi:10.1016/0370-2693(84)91161-4
  %%CITATION = doi:10.1016/0370-2693(84)91161-4;%%
  %98 citations counted in INSPIRE as of 08 Feb 2018

%\cite{Minkowski:1977sc}
\bibitem{Minkowski:1977sc} 
  P.~Minkowski,
  %``$\mu \to e\gamma$ at a Rate of One Out of $10^{9}$ Muon Decays?,''
  Phys.\ Lett.\  {\bf 67B}, 421 (1977).
%  doi:10.1016/0370-2693(77)90435-X
  %%CITATION = doi:10.1016/0370-2693(77)90435-X;%%
  %3000 citations counted in INSPIRE as of 08 Feb 2018

%\cite{Dudas:2009uq}
\bibitem{Dudas:2009uq} 
  E.~Dudas, Y.~Mambrini, S.~Pokorski and A.~Romagnoni,
  %``(In)visible Z-prime and dark matter,''
  JHEP {\bf 0908}, 014 (2009)
%  doi:10.1088/1126-6708/2009/08/014
  [arXiv:0904.1745 [hep-ph]].
  %%CITATION = doi:10.1088/1126-6708/2009/08/014;%%
  %101 citations counted in INSPIRE as of 08 Feb 2018

%\cite{Georgi:1974sy}
\bibitem{Georgi:1974sy} 
  H.~Georgi and S.~L.~Glashow,
  %``Unity of All Elementary Particle Forces,''
  Phys.\ Rev.\ Lett.\  {\bf 32}, 438 (1974).
%  doi:10.1103/PhysRevLett.32.438
  %%CITATION = doi:10.1103/PhysRevLett.32.438;%%
  %4482 citations counted in INSPIRE as of 26 Dec 2017

%\cite{Pati:1974yy}
\bibitem{Pati:1974yy} 
  J.~C.~Pati and A.~Salam,
  %``Lepton Number as the Fourth Color,''
  Phys.\ Rev.\ D {\bf 10}, 275 (1974)
  Erratum: [Phys.\ Rev.\ D {\bf 11}, 703 (1975)].
%  doi:10.1103/PhysRevD.10.275, 10.1103/PhysRevD.11.703.2
  %%CITATION = doi:10.1103/PhysRevD.10.275, 10.1103/PhysRevD.11.703.2;%%
  %4342 citations counted in INSPIRE as of 26 Dec 2017

%\cite{Georgi:1974my}
\bibitem{Georgi:1974my} 
  H.~Georgi,
  %``The State of the Art—Gauge Theories,''
  AIP Conf.\ Proc.\  {\bf 23}, 575 (1975);\\
%  doi:10.1063/1.2947450
  %%CITATION = doi:10.1063/1.2947450;%%
  %198 citations counted in INSPIRE as of 26 Dec 2017
%
%\cite{Fritzsch:1974nn}
%\bibitem{Fritzsch:1974nn} 
  H.~Fritzsch and P.~Minkowski,
  %``Unified Interactions of Leptons and Hadrons,''
  Annals Phys.\  {\bf 93}, 193 (1975).
%  doi:10.1016/0003-4916(75)90211-0
  %%CITATION = doi:10.1016/0003-4916(75)90211-0;%%
  %1620 citations counted in INSPIRE as of 26 Dec 2017

%\cite{Dimopoulos:1979es}
\bibitem{Dimopoulos:1979es} 
  S.~Dimopoulos and L.~Susskind,
  %``Mass Without Scalars,''
  Nucl.\ Phys.\ B {\bf 155}, 237 (1979).
%  doi:10.1016/0550-3213(79)90364-X
  %%CITATION = doi:10.1016/0550-3213(79)90364-X;%%
  %1175 citations counted in INSPIRE as of 26 Dec 2017
%\cite{Farhi:1980xs}

\bibitem{Farhi:1980xs} 
  E.~Farhi and L.~Susskind,
  %``Technicolor,''
  Phys.\ Rept.\  {\bf 74}, 277 (1981);
%  doi:10.1016/0370-1573(81)90173-3
  %%CITATION = doi:10.1016/0370-1573(81)90173-3;%%
  %1065 citations counted in INSPIRE as of 26 Dec 2017

%\cite{Agrawal:2011ze}
\bibitem{Agrawal:2011ze} 
  P.~Agrawal, S.~Blanchet, Z.~Chacko and C.~Kilic,
  %``Flavored Dark Matter, and Its Implications for Direct Detection and Colliders,''
  Phys.\ Rev.\ D {\bf 86}, 055002 (2012)
%  doi:10.1103/PhysRevD.86.055002
  [arXiv:1109.3516 [hep-ph]].
  %%CITATION = doi:10.1103/PhysRevD.86.055002;%%
  %89 citations counted in INSPIRE as of 08 Feb 2018

%\cite{Arcadi:2017kky}
\bibitem{Arcadi:2017kky} 
G.~Arcadi, M.~Dutra, P.~Ghosh, M.~Lindner, Y.~Mambrini, M.~Pierre, S.~Profumo and F.~S.~Queiroz,
%``The Waning of the WIMP? A Review of Models, Searches, and Constraints,''
arXiv:1703.07364 [hep-ph].
%%CITATION = ARXIV:1703.07364;%%
%75 citations counted in INSPIRE as of 17 Feb 2018

%\cite{Allahverdi:2017edd}
\bibitem{Allahverdi:2017edd} 
R.~Allahverdi, P.~S.~B.~Dev and B.~Dutta,
%``A Simple Testable Model of Baryon Number Violation: Baryogenesis, Dark Matter, Neutron-Antineutron Oscillation and Collider Signals,''
Phys.\ Lett.\ B {\bf 02}, 019 (2018)
%doi:10.1016/j.physletb.2018.02.019
[arXiv:1712.02713 [hep-ph]].
%%CITATION = doi:10.1016/j.physletb.2018.02.019;%%
%1 citations counted in INSPIRE as of 17 Feb 2018

%\cite{Fornal:2018eol}
\bibitem{Fornal:2018eol} 
B.~Fornal and B.~Grinstein,
%``Dark Matter Interpretation of the Neutron Decay Anomaly,''
Phys.\ Rev.\ Lett.\  {\bf 120}, no. 19, 191801 (2018)
%doi:10.1103/PhysRevLett.120.191801
[arXiv:1801.01124 [hep-ph]].
%%CITATION = doi:10.1103/PhysRevLett.120.191801;%%
%19 citations counted in INSPIRE as of 15 Jul 2018

%\cite{Garny:2018icg}
\bibitem{Garny:2018icg} 
M.~Garny, J.~Heisig, M.~Hufnagel and B.~Lülf,
%``Top-philic dark matter within and beyond the WIMP paradigm,''
Phys.\ Rev.\ D {\bf 97}, no. 7, 075002 (2018)
%doi:10.1103/PhysRevD.97.075002
[arXiv:1802.00814 [hep-ph]].
%%CITATION = doi:10.1103/PhysRevD.97.075002;%%
%7 citations counted in INSPIRE as of 15 Jul 2018

%\cite{Assad:2017iib}
\bibitem{Assad:2017iib} 
N.~Assad, B.~Fornal and B.~Grinstein,
%``Baryon Number and Lepton Universality Violation in Leptoquark and Diquark Models,''
Phys.\ Lett.\ B {\bf 777}, 324 (2018)
%doi:10.1016/j.physletb.2017.12.042
[arXiv:1708.06350 [hep-ph]].
%%CITATION = doi:10.1016/j.physletb.2017.12.042;%%
%20 citations counted in INSPIRE as of 16 Jul 2018

%\cite{DiLuzio:2017vat}
\bibitem{DiLuzio:2017vat} 
L.~Di Luzio, A.~Greljo and M.~Nardecchia,
%``Gauge leptoquark as the origin of B-physics anomalies,''
Phys.\ Rev.\ D {\bf 96}, no. 11, 115011 (2017)
%doi:10.1103/PhysRevD.96.115011
[arXiv:1708.08450 [hep-ph]].
%%CITATION = doi:10.1103/PhysRevD.96.115011;%%
%38 citations counted in INSPIRE as of 16 Jul 2018



%\cite{Barbieri:2015yvd}
\bibitem{Barbieri:2015yvd} 
R.~Barbieri, G.~Isidori, A.~Pattori and F.~Senia,
%``Anomalies in $B$-decays and $U(2)$ flavour symmetry,''
Eur.\ Phys.\ J.\ C {\bf 76}, no. 2, 67 (2016)
%  doi:10.1140/epjc/s10052-016-3905-3
[arXiv:1512.01560 [hep-ph]].
%%CITATION = doi:10.1140/epjc/s10052-016-3905-3;%%
%95 citations counted in INSPIRE as of 02 Feb 2018


%\cite{DiLuzio:2017chi}
\bibitem{DiLuzio:2017chi} 
L.~Di Luzio and M.~Nardecchia,
%``What is the scale of new physics behind the $B$-flavour anomalies?,''
Eur.\ Phys.\ J.\ C {\bf 77}, no. 8, 536 (2017)
%  doi:10.1140/epjc/s10052-017-5118-9
[arXiv:1706.01868 [hep-ph]].
%%CITATION = doi:10.1140/epjc/s10052-017-5118-9;%%
%12 citations counted in INSPIRE as of 02 Feb 2018

%\cite{Choudhury:2017qyt}
\bibitem{Choudhury:2017qyt} 
  D.~Choudhury, A.~Kundu, R.~Mandal and R.~Sinha,
  %``Minimal unified resolution to $R_{K^{(*)}}$ and $R(D^{(*)})$ anomalies with lepton mixing,''
  Phys.\ Rev.\ Lett.\  {\bf 119}, no. 15, 151801 (2017)
%  doi:10.1103/PhysRevLett.119.151801
  [arXiv:1706.08437 [hep-ph]].
  %%CITATION = doi:10.1103/PhysRevLett.119.151801;%%
  %15 citations counted in INSPIRE as of 04 May 2018

%\cite{Blanke:2018sro}
\bibitem{Blanke:2018sro} 
M.~Blanke and A.~Crivellin,
%``$B$ Meson Anomalies in a Pati-Salam Model within the Randall-Sundrum Background,''
Phys.\ Rev.\ Lett.\  {\bf 121}, no. 1, 011801 (2018)
%doi:10.1103/PhysRevLett.121.011801
[arXiv:1801.07256 [hep-ph]].
%%CITATION = doi:10.1103/PhysRevLett.121.011801;%%
%32 citations counted in INSPIRE as of 09 Sep 2018

%\cite{Crivellin:2018yvo}
\bibitem{Crivellin:2018yvo} 
A.~Crivellin, C.~Greub, F.~Saturnino and D.~Müller,
%``Importance of Loop Effects in Explaining the Accumulated Evidence for New Physics in B Decays with a Vector Leptoquark,''
arXiv:1807.02068 [hep-ph].
%%CITATION = ARXIV:1807.02068;%%
%3 citations counted in INSPIRE as of 09 Sep 2018

%\cite{Aprile:2018dbl}
\bibitem{Aprile:2018dbl}
E.~Aprile {\it et al.} [XENON Collaboration],
%``Dark Matter Search Results from a One Tonne$\times$Year Exposure of XENON1T,''
arXiv:1805.12562 [astro-ph.CO].
%%CITATION = ARXIV:1805.12562;%%
%20 citations counted in INSPIRE as of 04 Jul 2018


%\cite{Akerib:2016vxi}
\bibitem{Akerib:2016vxi} 
D.~S.~Akerib {\it et al.} [LUX Collaboration],
%``Results from a search for dark matter in the complete LUX exposure,''
Phys.\ Rev.\ Lett.\  {\bf 118}, no. 2, 021303 (2017)
%doi:10.1103/PhysRevLett.118.021303
[arXiv:1608.07648 [astro-ph.CO]].
%%CITATION = doi:10.1103/PhysRevLett.118.021303;%%
%537 citations counted in INSPIRE as of 17 Feb 2018

%\cite{Dorsner:2016wpm}
\bibitem{Dorsner:2016wpm} 
  I.~Doršner, S.~Fajfer, A.~Greljo, J.~F.~Kamenik and N.~Košnik,
  %``Physics of leptoquarks in precision experiments and at particle colliders,''
  Phys.\ Rept.\  {\bf 641}, 1 (2016)
%  doi:10.1016/j.physrep.2016.06.001
  [arXiv:1603.04993 [hep-ph]].
  %%CITATION = doi:10.1016/j.physrep.2016.06.001;%%
  %113 citations counted in INSPIRE as of 04 May 2018

%\cite{Garny:2012vt}
\bibitem{Garny:2012vt} 
M.~Garny, A.~Ibarra and D.~Tran,
%``Constraints on Hadronically Decaying Dark Matter,''
JCAP {\bf 1208}, 025 (2012)
%  doi:10.1088/1475-7516/2012/08/025
[arXiv:1205.6783 [hep-ph]].
%%CITATION = doi:10.1088/1475-7516/2012/08/025;%%
%31 citations counted in INSPIRE as of 07 Feb 2018

%\cite{Arcadi:2013aba}
\bibitem{Arcadi:2013aba} 
G.~Arcadi and L.~Covi,
%``Minimal Decaying Dark Matter and the LHC,''
JCAP {\bf 1308}, 005 (2013)
%doi:10.1088/1475-7516/2013/08/005
[arXiv:1305.6587 [hep-ph]].
%%CITATION = doi:10.1088/1475-7516/2013/08/005;%%
%19 citations counted in INSPIRE as of 08 Apr 2018

%\cite{Arcadi:2014dca}
\bibitem{Arcadi:2014dca} 
G.~Arcadi, L.~Covi and F.~Dradi,
%``3.55 keV line in Minimal Decaying Dark Matter scenarios,''
JCAP {\bf 1507}, no. 07, 023 (2015)
%doi:10.1088/1475-7516/2015/07/023
[arXiv:1412.6351 [hep-ph]].
%%CITATION = doi:10.1088/1475-7516/2015/07/023;%%
%17 citations counted in INSPIRE as of 08 Apr 2018

%\cite{Davidson:1993qk}
\bibitem{Davidson:1993qk} 
S.~Davidson, D.~C.~Bailey and B.~A.~Campbell,
%``Model independent constraints on leptoquarks from rare processes,''
Z.\ Phys.\ C {\bf 61}, 613 (1994)
%doi:10.1007/BF01552629
[hep-ph/9309310].
%%CITATION = doi:10.1007/BF01552629;%%
%405 citations counted in INSPIRE as of 18 Feb 2018

%\cite{Valencia:1994cj}
\bibitem{Valencia:1994cj} 
G.~Valencia and S.~Willenbrock,
%``Quark - lepton unification and rare meson decays,''
Phys.\ Rev.\ D {\bf 50}, 6843 (1994)
%doi:10.1103/PhysRevD.50.6843
[hep-ph/9409201].
%%CITATION = doi:10.1103/PhysRevD.50.6843;%%
%67 citations counted in INSPIRE as of 18 Feb 2018

%\cite{Ade:2015xua}
\bibitem{Ade:2015xua} 
P.~A.~R.~Ade {\it et al.} [Planck Collaboration],
%``Planck 2015 results. XIII. Cosmological parameters,''
Astron.\ Astrophys.\  {\bf 594}, A13 (2016)
%  doi:10.1051/0004-6361/201525830
[arXiv:1502.01589 [astro-ph.CO]].
%%CITATION = doi:10.1051/0004-6361/201525830;%%
%3325 citations counted in INSPIRE as of 16 May 2017


%\cite{Agrawal:2010fh}
\bibitem{Agrawal:2010fh} 
  P.~Agrawal, Z.~Chacko, C.~Kilic and R.~K.~Mishra,
  %``A Classification of Dark Matter Candidates with Primarily Spin-Dependent Interactions with Matter,''
  arXiv:1003.1912 [hep-ph].
  %%CITATION = ARXIV:1003.1912;%%
  %80 citations counted in INSPIRE as of 02 Feb 2018
 
%%\cite{Aad:2011ch}
%\bibitem{Aad:2011ch} 
%G.~Aad {\it et al.} [ATLAS Collaboration],
%%``Search for first generation scalar leptoquarks in $pp$ collisions at $\sqrt{s}=7$ TeV with the ATLAS detector,''
%Phys.\ Lett.\ B {\bf 709}, 158 (2012)
%Erratum: [Phys.\ Lett.\ B {\bf 711}, 442 (2012)]
%%  doi:10.1016/j.physletb.2012.03.023, 10.1016/j.physletb.2012.02.004
%[arXiv:1112.4828 [hep-ex]].
%%%CITATION = doi:10.1016/j.physletb.2012.03.023, 10.1016/j.physletb.2012.02.004;%%
%%79 citations counted in INSPIRE as of 02 Feb 2018

%\cite{Mallot:1999qb}
\bibitem{Mallot:1999qb} 
  G.~K.~Mallot,
  %``The Spin structure of the nucleon,''
  Int.\ J.\ Mod.\ Phys.\ A {\bf 15S1}, 521 (2000)
  [eConf C {\bf 990809}, 521 (2000)]
%  doi:10.1142/S0217751X00005309
  [hep-ex/9912040].
  %%CITATION = doi:10.1142/S0217751X00005309;%%
  %34 citations counted in INSPIRE as of 03 Apr 2018



%\cite{Ellis:2000ds}
\bibitem{Ellis:2000ds} 
  J.~R.~Ellis, A.~Ferstl and K.~A.~Olive,
  %``Reevaluation of the elastic scattering of supersymmetric dark matter,''
  Phys.\ Lett.\ B {\bf 481}, 304 (2000)
%  doi:10.1016/S0370-2693(00)00459-7
  [hep-ph/0001005].
  %%CITATION = doi:10.1016/S0370-2693(00)00459-7;%%
  %355 citations counted in INSPIRE as of 03 Apr 2018

%\cite{Arcadi:2014tsa}
\bibitem{Arcadi:2014tsa} 
G.~Arcadi, L.~Covi and F.~Dradi,
%``LHC prospects for minimal decaying Dark Matter,''
JCAP {\bf 1410}, no. 10, 063 (2014)
%doi:10.1088/1475-7516/2014/10/063
[arXiv:1408.1005 [hep-ph]].
%%CITATION = doi:10.1088/1475-7516/2014/10/063;%%
%12 citations counted in INSPIRE as of 14 Apr 2018

%\cite{Aad:2015caa}
\bibitem{Aad:2015caa} 
G.~Aad {\it et al.} [ATLAS Collaboration],
%``Searches for scalar leptoquarks in pp collisions at $\sqrt{s}$ = 8 TeV with the ATLAS detector,''
Eur.\ Phys.\ J.\ C {\bf 76}, no. 1, 5 (2016)
%  doi:10.1140/epjc/s10052-015-3823-9
[arXiv:1508.04735 [hep-ex]].
%%CITATION = doi:10.1140/epjc/s10052-015-3823-9;%%
%84 citations counted in INSPIRE as of 02 Feb 2018


%\cite{Sirunyan:2017yrk}
\bibitem{Sirunyan:2017yrk} 
A.~M.~Sirunyan {\it et al.} [CMS Collaboration],
%``Search for third-generation scalar leptoquarks and heavy right-handed neutrinos in final states with two tau leptons and two jets in proton-proton collisions at $ \sqrt{s}=13 $ TeV,''
JHEP {\bf 1707}, 121 (2017)
%  doi:10.1007/JHEP07(2017)121
[arXiv:1703.03995 [hep-ex]].
%%CITATION = doi:10.1007/JHEP07(2017)121;%%
%25 citations counted in INSPIRE as of 02 Feb 2018

%\cite{Sirunyan:2018kzh}
\bibitem{Sirunyan:2018kzh} 
A.~M.~Sirunyan {\it et al.} [CMS Collaboration],
%``Constraints on models of scalar and vector leptoquarks decaying to a quark and a neutrino at $\sqrt{s}=$ 13 TeV,''
arXiv:1805.10228 [hep-ex].
%%CITATION = ARXIV:1805.10228;%%
%1 citations counted in INSPIRE as of 05 Jun 2018

%\cite{Khachatryan:2014rra}
\bibitem{Khachatryan:2014rra} 
  V.~Khachatryan {\it et al.} [CMS Collaboration],
  %``Search for dark matter, extra dimensions, and unparticles in monojet events in proton–proton collisions at $\sqrt{s} = 8$ TeV,''
  Eur.\ Phys.\ J.\ C {\bf 75}, no. 5, 235 (2015)
%  doi:10.1140/epjc/s10052-015-3451-4
  [arXiv:1408.3583 [hep-ex]].
  %%CITATION = doi:10.1140/epjc/s10052-015-3451-4;%%
  %381 citations counted in INSPIRE as of 02 Apr 2018

%\cite{Aad:2015zva}
\bibitem{Aad:2015zva} 
  G.~Aad {\it et al.} [ATLAS Collaboration],
  %``Search for new phenomena in final states with an energetic jet and large missing transverse momentum in pp collisions at $\sqrt{s}=$8 TeV with the ATLAS detector,''
  Eur.\ Phys.\ J.\ C {\bf 75}, no. 7, 299 (2015)
  Erratum: [Eur.\ Phys.\ J.\ C {\bf 75}, no. 9, 408 (2015)]
%  doi:10.1140/epjc/s10052-015-3517-3, 10.1140/epjc/s10052-015-3639-7
  [arXiv:1502.01518 [hep-ex]].
  %%CITATION = doi:10.1140/epjc/s10052-015-3517-3, 10.1140/epjc/s10052-015-3639-7;%%
  %283 citations counted in INSPIRE as of 02 Apr 2018

%\cite{Aaboud:2016tnv}
\bibitem{Aaboud:2016tnv} 
  M.~Aaboud {\it et al.} [ATLAS Collaboration],
  %``Search for new phenomena in final states with an energetic jet and large missing transverse momentum in $pp$ collisions at $\sqrt{s}=13$  TeV using the ATLAS detector,''
  Phys.\ Rev.\ D {\bf 94}, no. 3, 032005 (2016)
%  doi:10.1103/PhysRevD.94.032005
  [arXiv:1604.07773 [hep-ex]].
  %%CITATION = doi:10.1103/PhysRevD.94.032005;%%
  %194 citations counted in INSPIRE as of 02 Apr 2018

%\cite{CMS:2017tbk}
\bibitem{CMS:2017tbk} 
  CMS Collaboration [CMS Collaboration],
  %``Search for new physics in final states with an energetic jet or a hadronically decaying W or Z boson using $35.9~\mathrm{fb}^{-1}$ of data at $\sqrt{s} = 13~\mathrm{TeV}$,''
  CMS-PAS-EXO-16-048.
  %%CITATION = CMS-PAS-EXO-16-048;%%
  %14 citations counted in INSPIRE as of 02 Apr 2018

%\cite{Aaboud:2017phn}
\bibitem{Aaboud:2017phn} 
  M.~Aaboud {\it et al.} [ATLAS Collaboration],
  %``Search for dark matter and other new phenomena in events with an energetic jet and large missing transverse momentum using the ATLAS detector,''
  JHEP {\bf 1801}, 126 (2018)
%  doi:10.1007/JHEP01(2018)126
  [arXiv:1711.03301 [hep-ex]].
  %%CITATION = doi:10.1007/JHEP01(2018)126;%%
  %24 citations counted in INSPIRE as of 02 Apr 2018


%\cite{Alwall:2014hca}
\bibitem{Alwall:2014hca} 
  J.~Alwall {\it et al.},
  %``The automated computation of tree-level and next-to-leading order differential cross sections, and their matching to parton shower simulations,''
  JHEP {\bf 1407}, 079 (2014)
%  doi:10.1007/JHEP07(2014)079
  [arXiv:1405.0301 [hep-ph]].
  %%CITATION = doi:10.1007/JHEP07(2014)079;%%
  %2652 citations counted in INSPIRE as of 03 Apr 2018

%\cite{Alloul:2013bka}
\bibitem{Alloul:2013bka} 
  A.~Alloul, N.~D.~Christensen, C.~Degrande, C.~Duhr and B.~Fuks,
  %``FeynRules  2.0 - A complete toolbox for tree-level phenomenology,''
  Comput.\ Phys.\ Commun.\  {\bf 185}, 2250 (2014)
%  doi:10.1016/j.cpc.2014.04.012
  [arXiv:1310.1921 [hep-ph]].
  %%CITATION = doi:10.1016/j.cpc.2014.04.012;%%
  %822 citations counted in INSPIRE as of 03 Apr 2018



%\cite{Ball:2012cx}
\bibitem{Ball:2012cx} 
  R.~D.~Ball {\it et al.},
  %``Parton distributions with LHC data,''
  Nucl.\ Phys.\ B {\bf 867}, 244 (2013)
%  doi:10.1016/j.nuclphysb.2012.10.003
  [arXiv:1207.1303 [hep-ph]].
  %%CITATION = doi:10.1016/j.nuclphysb.2012.10.003;%%
  %1058 citations counted in INSPIRE as of 03 Apr 2018

%\cite{An:2013xka}
\bibitem{An:2013xka} 
  H.~An, L.~T.~Wang and H.~Zhang,
  %``Dark matter with $t$-channel mediator: a simple step beyond contact interaction,''
  Phys.\ Rev.\ D {\bf 89}, no. 11, 115014 (2014)
%  doi:10.1103/PhysRevD.89.115014
  [arXiv:1308.0592 [hep-ph]].
  %%CITATION = doi:10.1103/PhysRevD.89.115014;%%
  %102 citations counted in INSPIRE as of 03 Apr 2018


%\cite{Aguilar:2016kjl}
\bibitem{Aguilar:2016kjl} 
  M.~Aguilar {\it et al.} [AMS Collaboration],
  %``Antiproton Flux, Antiproton-to-Proton Flux Ratio, and Properties of Elementary Particle Fluxes in Primary Cosmic Rays Measured with the Alpha Magnetic Spectrometer on the International Space Station,''
  Phys.\ Rev.\ Lett.\  {\bf 117}, no. 9, 091103 (2016).
%  doi:10.1103/PhysRevLett.117.091103
  %%CITATION = doi:10.1103/PhysRevLett.117.091103;%%
  %114 citations counted in INSPIRE as of 07 Feb 2018

%\cite{Cuoco:2017iax}
\bibitem{Cuoco:2017iax} 
  A.~Cuoco, J.~Heisig, M.~Korsmeier and M.~Krämer,
  %``Constraining heavy dark matter with cosmic-ray antiprotons,''
  arXiv:1711.05274 [hep-ph].
  %%CITATION = ARXIV:1711.05274;%%
  %5 citations counted in INSPIRE as of 07 Feb 2018



%\cite{Akerib:2016lao}
\bibitem{Akerib:2016lao} 
D.~S.~Akerib {\it et al.} [LUX Collaboration],
%``Results on the Spin-Dependent Scattering of Weakly Interacting Massive Particles on Nucleons from the Run 3 Data of the LUX Experiment,''
Phys.\ Rev.\ Lett.\  {\bf 116}, no. 16, 161302 (2016)
%  doi:10.1103/PhysRevLett.116.161302
[arXiv:1602.03489 [hep-ex]].
%%CITATION = doi:10.1103/PhysRevLett.116.161302;%%
%114 citations counted in INSPIRE as of 02 Feb 2018

%\cite{Akerib:2015cja}
\bibitem{Akerib:2015cja} 
D.~S.~Akerib {\it et al.} [LZ Collaboration],
%``LUX-ZEPLIN (LZ) Conceptual Design Report,''
arXiv:1509.02910 [physics.ins-det].
%%CITATION = ARXIV:1509.02910;%%
%185 citations counted in INSPIRE as of 02 Feb 2018

%\cite{Patrignani:2016xqp}
\bibitem{Patrignani:2016xqp} 
C.~Patrignani {\it et al.} [Particle Data Group],
%``Review of Particle Physics,''
Chin.\ Phys.\ C {\bf 40}, no. 10, 100001 (2016).
%doi:10.1088/1674-1137/40/10/100001
%%CITATION = doi:10.1088/1674-1137/40/10/100001;%%
%3526 citations counted in INSPIRE as of 05 Jun 2018

\end{thebibliography}
\end{document}